\documentstyle[prb,preprint,aps]{revtex}

\def\prb{Phys. Rev. B }

\begin{document}
\title{ Onset voltage shift due to non-zero Landau ground state level in coherent magnetotransport }
\author{Dae Kwan Kim\footnote{E-Mail:kimd@ee.eng.ohio$-$state.edu} and 
Patrick Roblin\footnote{E-Mail:roblin@ee.eng.ohio$-$state.edu} }
\address{Department of Electrical Engineering, The Ohio State University, Columbus, Ohio 43210}
\maketitle
\vspace{0.5cm}
\begin{abstract}
Coherent electron transport in double-barrier heterostructures with parallel 
electric and magnetic fields is analyzed theoretically and 
with the aid of a quantum simulator accounting for 3-dimensional transport effects. 
The onset-voltage shift induced by the magnetic field in
resonant tunneling diodes, which was previously 
attributed to the cyclotron frequency $w_c$ inside the well is 
found to arise from
an upward shift of the non-zero ground (lowest) Landau state energy in 
the entire quantum region where coherent transport takes place. 
The spatial dependence of the cyclotron frequency is accounted for and verified
to have a negligible impact on resonant tunneling
for the device and magnetic field strength considered.
A correction term for the onset-voltage shift arising from the 
magnetic field dependence of the chemical potential is also derived.
The Landau ground state with its nonvanishing finite harmonic oscillator energy
$ \hbar w_c /2$ is verified however to be the principal contributor to 
the onset voltage shift at low temperatures.
\end{abstract}
\draft
\newpage

\section{Introduction}
One of the most fundamental results of quantum theory is the fact that the
harmonic oscillator has a ground state with a  {\it non-zero} energy $ \hbar w_c /2$. 
This energy is usually ignored in most field calculations because it is merely a constant and 
it is therefore treated as a reference point in energy. 

The quantum state of an electron in a constant and uniform magnetic field 
is described by a {\it harmonic oscillator} with cyclotron frequency $w_c$. 
Indeed from a classical mechanics point of view, 
this results from the fact that the electrons move through the
device according to helicoidal trajectories,
and that the projected motion along an arbitrary transverse
axis is that of a simple harmonic oscillator. 
Varying the magnetic field changes the cyclotron frequency
and is found in turns to induce a {\it shift of the onset voltage} 
in the IV characteristic of double barrier heterostructures.
We demonstrate in this paper that this 
collective phenomenon in the electron current occurs due to the
varying energy value $ \hbar w_c /2$ of  the ground-state of the 
harmonic-oscillator representing the electrons (see Fig.\ \ref{vshif}).   

The famous Casimir effect \cite{cas} has the same origin in nature.
Each mode of the electromagnetic field is represented by a harmonic oscillator.
The finite ground state energy of an electromagnetic field between parallel metal plates depends on the separation of the plates, which is similar 
to the dependence upon the  varying magnetic field in the system
studied here.
The Casimir effect which was predicted in 1948 was recently verified experimentally by Lamoreaux in 1996 \cite{lar}.

Other known examples are the Lamb shift \cite{lam}, and the unfreezing of 
liquid helium at extremely low temperatures. Nevertheless
phenomena which have a direct relation with the nonvanishing, finite ground-state (zero-point) energy of the corresponding harmonic oscillator are very {\it few}. In this paper one more occurrence is demonstrated for magnetotransport in semiconductor heterostructures. 

The magnetotransport phenomenon in multiple barrier heterostructures with a 
magnetic field applied parallel to the electric field was first investigated 
and reported in the pioneering work of Mendez, Esaki and Wang \cite{mend}.   
When both the magnetic field and electric field are perpendicular to 
the heterostructure interface \cite{chan}, the electron motion becomes 
quasi-one dimensional \cite{kelly}, as the electron states in the plane 
perpendicular to the layered direction are quantized in discrete Landau levels. 
This dimensional reduction of the electron motion has a profound effect on 
the transmission probability and current-voltage characteristic when compared 
with the case of zero magnetic field.   

The experimental work on GaAs/GaAlAs double barrier structures incorporating both a parallel  electric field and a magnetic field showed some remarkable features,
including a shift of onset voltage and enhanced peak-to-valley-ratio \cite{mend,lead}. However, these observations have not been fully accounted for from a theoretical standpoint, and such a theoretical analysis is needed to understand the underlying physical processes and inspire possible device applications.

Specifically in Ref. 4 the shift of onset voltage  with 
increasing  magnetic fields was predicted to be given by the energy
$\hbar w_c /2$ associated with the cyclotron frequency
inside the well region. 
In this paper we demonstrate instead 
that the onset voltage shift originates from 
an upward shift of the non-zero ground (lowest) Landau state energy in 
the perpendicular plane throughout the {\it entire} region of the quantum 
structure where coherent transport takes place, 
including the left contact, rather than only inside the well. 
We also derive and numerically calculate a correction term for the
onset voltage shift which arises from the magnetic field dependence of the
chemical potential.
Finally we have verified using our 3D transport model that the 
spatial dependence of the cyclotron frequency 
has a negligible impact on resonant tunneling
in the test device for the magnetic field strength considered.

From a fundamental physical point of view, the fact that the {\it nonvanishing} 
ground state energy $1/2 \cdot \hbar w_c$ reveals itself in a shift of the 
onset voltage is a very noticeable phenomenon.  
We account here for this abnormal feature with the aid of a 
quantum simulator {\bf based on a 3-dimensional transport picture}
which incorporates the Landau states in the perpendicular plane.  

Undoped Al$_{0.3}$Ga$_{0.7}$As/GaAs/Al$_{0.3}$Ga$_{0.7}$As double barrier structures with heavily doped $n^+$-GaAs left and right contact layers are considered. The donor doping density is $1 \times 10^{18}/cm^3$.
The conduction-band profile without an applied voltage is shown in the inset of Fig.\ \ref{tvef}.
The effective mass of the electrons for GaAs and Al$_{0.3}$Ga$_{0.7}$As are given by $m^*=0.067m, 0.092m$, respectively, where $m$ is the electron rest-mass.

 Two structures with a barrier and a well width of $34 \AA$ (structure  A) and $50 \AA$ (structure B) respectively are examined.
We assume that the donors in the contacts are completely ionized ($n=N_d$) due to the heavy doping density.
A tight-binding band in the superlattice direction and a parabolic band in the perpendicular plane are used.  
Spin splitting in a magnetic field is ignored for simplicity.
Finally we focus on the low temperature case, $T=4.2K$.

\section{Theory}
The electron Hamiltonian in the heterostructure in the most general case includes interaction terms which describe disorder scattering by phonons, impurities, interface roughness, etc. Since the electron mean free path at very low temperatures is relatively large compared to the heterostructure size under investigation ($w_c \tau \gg 1$),
we can safely neglect scattering. Therefore, only the coherent, ballistic electron motion will be considered here and the Schr\"{o}dinger equation to be solved is 
\begin{equation}
H_e |\Psi \rangle =i \hbar \frac{\mbox{d}}{\mbox{dt}} |\Psi \rangle.
\end{equation}

In the plane perpendicular to the superlattice direction, the electrons move classically in a circular path. In a quantum treatment the electron energy is quantized in Landau states \cite{lan},  which are given by
 the quantized harmonic oscillator states,
\begin{equation}
E_{p} =(L+1/2 )\hbar w_c(n), 
\end{equation}
where $w_c(n)= eB/{m^{*}(n)}$ is the cyclotron frequency in the magnetic field $B$ at the lattice site $n$ in the superlattice direction.
In the following discussion we will see that the lowest Landau state $L=0$ is not just a reference point but gives an important contribution to the current flow in the double barrier heterostructure.

We expand the electron wave function in a basis consisting of generalized Wannier functions in the superlattice direction ($x$-direction) \cite{rob} and of the Landau states in the $(y, z)$ plane which result when
the  Landau gauge is chosen for the vector potential of the magnetic field,
\begin{equation}
|\Psi\rangle=\sum_{n,L}\int f(k_y,L,n) |k_y,L,n\rangle dk_y,
\end{equation}
where $f(k_y,L,n)$ is the envelope function of the electron states.
The matrix element for the heterostructure Hamiltonian between 
generalized Wannier-Landau states is \cite{rob}
\begin{equation}
 \langle  k_y,L,n| H_{e}| k^{\prime}_y ,L^{\prime},n^{\prime} \rangle
=H_{e,nn^{\prime}}(L)\delta_{L,L^{\prime}}
\delta (k_y -k^{\prime}_y),
\end{equation}
where
\begin{equation}
H_{e,nn^{\prime}}(L) = H_{nn^{\prime}}+
(L+1/2 )\hbar w_c(n)\delta_{nn^{\prime}},
\end{equation}
and
\begin{eqnarray}
H_{nn^{\prime}} &=&-\frac{\hbar^2}{2a^2\sqrt{m^{*}(n)
 m^{*}(n+1)} }\delta_{n+1,n^{\prime}} \nonumber \\
& &-\frac{\hbar^2}{2a^2\sqrt{m^{*}(n) m^{*}(n-1)} }\delta_{n-1,n^{\prime}}
+\left[ \frac{\hbar^2}{a^2 m^{*} (n)}+
E_{con}(n)-eV_{app}(n)  \right]\delta_{n,n^{\prime}}.
\end{eqnarray}
Here $E_{con}(n)$ is the conduction band edge and $V_{app}(n)$ is the electrostatic potential at the lattice site $n$.

The transversal mass in the perpendicular part of the Hamiltonian matrix also varies
 along the longitudinal direction. We introduce a new Hamiltonian matrix 
 which absorbs this $n$-dependent transversal part as
\begin{equation}
H_{e,nn^{\prime}}(L) =
 \tilde{H}_{nn^{\prime}}(L)+
(L+1/2 )\hbar w_c(0)\delta_{nn^{\prime}},
\end{equation}
where
\begin{equation}
\tilde{H}_{nn^{\prime}}(L) = H_{nn^{\prime}}+
(L+1/2 ) \hbar w_c(0) [m^*(0)/m^*(n)-1]\delta_{nn^{\prime}}.
\label{equ8}
\end{equation}
Here $\tilde{H}_{nn^{\prime}}$ depends on the perpendicular states L, 
incorporating
3-dimensional {\bf band-structure} effects in the longitudinal motion of the electrons.

Thus the Schr\"{o}dinger equation reduces to the following $3$-dimensional envelope equation for $f(k_y,L,n)$:
\begin{eqnarray}
i \hbar \frac{ \mbox{d}f(k_y,L,n) }{ \mbox{dt}} &=&(L+1/2 )\hbar w_{c0} f(k_y,L,n)+
\sum_{m}  \tilde{H}_{nm} f(k_y,L,m),
\end{eqnarray}
where $w_{c0}=eB/{m^{*}(0)}$ is the cyclotron frequency at the left contact.

The incident electrons in the left contact occupy the sub-bands $E_{L_0}=E_{0x}+(L_0+1/2 )\hbar w_{c0}$, where $E_{0x}=\hbar^2 k_x ^2 /(2m^*(0))$.
Since in the absence of scattering an incident ballistic electron conserves its Landau level $L_0$ and its total energy $E_{L_0}$, as it goes 
through the structure, 
the envelope function  solution of Eq.(9) in Landau gauge takes the form
\begin{eqnarray}
f(k_y,L,n)&=&e^{-iE_{L_0} t} f_0(n)\delta (k_y-k_{y0}) \delta_{L L_0}.
\end{eqnarray}
When Eq.(10) is substituted into the Schr\"{o}dinger equation (9), we arrive at the time independent form
\begin{eqnarray}
E_{0x}f_0(n) &=& \sum_{m}  \tilde{H}_{nm} f_0(m)
\end{eqnarray}
Note that the longitudinal electron wave function $f_0(n)$ depends on the perpendicular 
Landau sub-band level $L_0$, since the second term in the right hand side of Eq.(8) 
is a function of the Landau level. This is a 3-dimensional transport effect which 
was exactly implemented in our numerical analysis.
  
The electron currents for an incident and transmitted state $(E_{0x},L_0)$ are respectively,
\begin{eqnarray}
j_{I0}(E_{0x},L_0) &=&e v_L(E_{0x}) \nonumber \\
j_{T0}(E_{0x},L_0) &=&e |f_0(n_R)| ^2 v_R(E_{0x}),
\end{eqnarray}
where $v_L,v_R$ are the electron velocities at the left and right contacts, and $n_R$ is
the lattice-site index of the right contact.
The transmission probability  from left to right contact is then defined as
\begin{equation}
T_{L\rightarrow R}(E_{0x},L_0)= j_{T0}(E_{0x},L_0)/j_{I0}(E_{0x},L_0). 
\end{equation}
It is noted that the longitudinal transmission probability depends on the perpendicular state $L_0$. Fig.\ \ref{tvef} shows the transmission  probability for different applied biases.

In order to calculate the total diode current, the chemical potential corresponding to a doping density $n$ in
the contacts
in the presence of a magnetic field should be determined. 
The chemical potential has an oscillatory dependence on the magnetic field which is determined from the carrier density through the following integration relation, 
\begin{equation}
n=N_d = (\sqrt{2 m^*}/\pi \hbar)D
 \sum_{L_0=0}^{\infty}\int_{0}^{\infty}
(1/\sqrt{E_{0x}}) f_{FD}(E_{0x},L_0,\mu(B)) \mbox{d}E_{0x},
\end{equation}
where $D=eB/h$ is a degeneracy factor in the plane perpendicular to the superlattice direction and where the
Fermi-Dirac function $f_{FD}$ is given by  
\begin{equation}
f_{FD}(E_{0x},L_0,\mu)=\{ \exp[(E_{0x}+(L_0+1/2)\hbar w_{c0} -\mu)/kT]+1 \}^{-1}.
\end{equation}
The chemical potential $\mu$ is a complex function of the electron density $n$, magnetic field and temperature.
Its numerical calculation shows that the chemical potential $\mu$ oscillates by a large quantity when the magnetic field $B$ is strong; for example, there is a difference of $4.4$ meV between its value at $B=13$ and at $B=17$ Telsa.
It is important to account for this behavior to obtain an accurate value of the current. 

Under strong magnetic fields, the density of electronic states (see Eq.(14)) 
drastically departs from the magnetic-field free case
and in addition
an upward shift $\hbar w_c(n) /2$ of electron's lateral energy with respect 
to the conduction band minimum occurs in the entire quantum region.
As we will see below, the current integration over 
electronic states in the local emitter will strongly depend on the
$\hbar w_{c0}/2$ shift
via the Fermi-Dirac distribution function, Eq.(15). This will
lead in turn to the onset voltage shift in the current-voltage curve.

The total diode current density is then 
\begin{equation}
I_{tot}= 
(e D/\pi \hbar) \sum_{L_0=0}^{\infty}\int T_{L\rightarrow R}(E_{0x},L_0) 
\cdot f_{FD}(E_{0x},L_0,\mu(B))\mbox{d}E_{0x}-[L \leftrightarrow R],
\end{equation}
where 
the second term in the right-hand side is the backward contribution to the total current. 
In understanding the abnormal characteristics of the current, it is helpful to 
define an effective chemical potential $\mu_{eff}(B,L_0)$ dependent 
upon the Landau level $L_0$ as follows
\begin{equation}
\mu_{eff}(B,L_0)=\mu(B)-(2L_0+1)\hbar w_{c0}/2
\end{equation}
At extremely low temperatures such as 4.2 K, the Fermi-Dirac function becomes flat for energies below the chemical potential.
To contribute to the total current, the longitudinal energy of electron 
states $(E_{0x},L_0)$ should satisfy $E_{0x} \le \mu_{eff}(B,L_0)$.

As is indicated by Eq.(16) the
resonant tunneling current through a double barrier structure generally depends on the details 
of the transmission probability, the density of electronic states (DOS) and
the chemical potential $\mu$ (or Fermi energy). Among these, the DOS is much more affected by strong magnetic fields 
rather than the other properties.
As the summation and factor $1/\sqrt{E_{0x}}$ in Eq.(14) indicate, the DOS severely changes due to the quantization 
of the lateral energy of electrons in the left (right) local contact.
The electron states are then classified using the longitudinal sub-band index $L_0$. 
At the temperature of 4.2 K, the  longitudinal energy range depends on $L_0$ according 
to $E_{0x} \le \mu_{eff} (B,L_0)$.  Thus the effective 
chemical potential $\mu_{eff} (B,L_0)$ for a particular sub-band $L_0$ is defined to 
indicate the cutoff obtained in the longitudinal energy space.

Under such conditions, the current for an occupied sub-band $L_0$ reduces to
\begin{equation}
I_{tot,L_0} \approx
(e D/\pi \hbar)\int _{0}^{\mu_{eff}}
T_{L\rightarrow R}(E_{0x},L_0)\mbox{d}E_{0x} -[L \leftrightarrow R],
\end{equation}
where the minimum of the conduction band at the left contact has been set to zero.
Therefore, the area of the transmission probability determines the contribution of each occupied sub-band to the total current.
Fig.\ \ref{ivst} shows the currents calculated for structure A and B. 
As the voltage is increased, the lowest $L_0$ sub-band contributes to the current first with higher Landau levels contributing at higher voltages. 

\section{Results}
We note  here two important features in the results obtained.
First, a shift of the onset voltage depending on the magnetic field strength is observed \cite{mend}. It is sharper for the structure with thicker barriers.
Second, the current peaks are enhanced in regions of strong fields in both structures \cite{lead}. For a thicker barrier (structure B) this current increase is quite remarkable over a wide range of $B$ field. 

Let us analyze why these features occur in the presence of a magnetic field.
As the applied voltage increases, the quasi-resonant energy $E_r$ (see the inset of 
Fig.\ \ref{tvef}) is lowered.
At the same time, the conduction band profile has a decreased symmetry due to its tilt. It results that the height and width of the peak transmission probability near $E_r$ decreases, as can be verified in Fig.\ \ref{tvef}.
The energy width of the transmission probability is narrower in RTD structures with a thicker barrier. Therefore the transmission probability is  much sharper for structure B than for structure A.
These properties explain the increased sharpness of the current threshold and 
the downward decline of the current plateau in structure B (see  Fig.\ \ref{ivst}).

Let us consider the lowest Landau sub-band $L_0=0$, whose effective chemical
potential is $\mu_{eff}(B,0)=\mu(B)-\hbar w_{c0}(B)/2$.
In the absence of  a magnetic field, 
the onset voltage for RTD diodes with symmetric barriers and for 
$eV_D=E_{f,L}-E_{f,R}$ positive
is well-known to be approximately given by
\begin{equation}
V_{D,on}/2 = E_r -E_{f,L} = E_r - \mu(0),
\end{equation}
where $E_{f,L}$ is the Fermi energy level for the electron carriers on the left contact.
In the presence of a magnetic field $\mu(0)$ is just replaced by $\mu_{eff}(B,0)$
and $E_r$ can be assumed to remain unchanged.
Indeed the dependence of $E_r$ upon the magnetic field $B$ which arises from the
variation of the mass along the superlattice direction (see Eq.(8)) was verified to be negligible
in the structures studied.
As a result  the shift $\Delta V_{D,on}$  of the onset voltage is approximately given by
\begin{equation}
 \Delta V_{D,on}/2 =[V_{D,on}(B) -V_{D,on}(0)]/2   
 \simeq [\mu(0)-\mu(B)]+\hbar w_{c0}(B)/2,
\end{equation}
which is equal to the difference in effective chemical potential.
Fig.\ \ref{vshif} indicates that the dominant contribution to the onset voltage shift originates from 
the non-zero energy $\hbar w_{c0}/2$ of the Landau ground (lowest) state which is a unique feature.
Note also that the superposed oscillation in $\Delta V_{D,on}$ 
has the same origin, namely the oscillation of the chemical potential,
as in other phenomena such as de Haas-van Alphen effect \cite{zim}.

Finally let us look at the increase of the current peak (see Fig.\ \ref{ivst}).
The current density Eq.(18) has a weight factor proportional to the degeneracy $D$.
As discussed, the sharpness of the transmission coefficient depends on the structure parameters such as the barrier widths as well as the  conduction-band barrier heights. The enhancement of the current peaks in presence of magnetic field depends on the width of the transmission coefficient peak  near the quasi-resonant energy $E_r$.
Therefore, a structure with a sharper transmission probability has a more enhanced peak.

The RTD diode current in Fig.\ \ref{ivst} for structure B increases step by step  for a weaker magnetic field, with plateaus slowly declining downward.  
However for a strong magnetic field at  $B=20$ Telsa where the two lowest Landau levels of the emitter are occupied, its diode current rises to its maximum peak value in just one step.
For structure A (B) in a strong field of $B=20$ Telsa, an enhancement of the current peak by $10 (37) \; \%$ is observed.

\section{Conclusion}
In this paper we have analyzed
coherent electron transport in double-barrier heterostructures 
with parallel electric and magnetic fields
{\bf with the aid of a quantum simulator accounting for 3-dimensional transport}.
The presence of a magnetic field affect severely 
the electron transport in a resonant tunneling diode causing 
a shift of the onset voltage.

Our analysis establishes that
the onset-voltage shift induced by the magnetic field arises from
an upward shift of the non-zero ground (lowest) Landau state energy in 
the {\it entire} quantum region where coherent transport takes place.
Note that quantum region therefore includes the left contact rather than only 
the well as previously reported \cite{mend}.  Actually in this work the 
left contact is used as the cyclotron energy reference.
Indeed our 3-dimensional {\bf transport}  analysis accounts for
the spatial dependence of the cyclotron frequency which is, however, verified
to have a relatively small impact on resonant tunneling
in the device analyzed for the magnetic field strength considered.
A larger correction term for the onset-voltage shift arising from the 
magnetic field dependence of the chemical potential is also derived.

The comprehensive 3D analysis reported in this paper reveals that
the Landau ground state with its nonvanishing finite harmonic oscillator energy
$ \hbar w_c(0) /2$ is indeed the principal contributor to 
the onset voltage shift at low temperatures.
Finally it is also shown in this paper how
the abnormal increase of the current peak depends
on the external magnetic field as well as the structure parameters
of the double barrier heterostructure.

 \vspace{0.5cm}
\noindent
{\large{\bf Acknowledgements}}
\vspace{0.5cm} \\

This work was supported in part by the Korea Science and Engineering Foundation (KOSEF). 
The first author is grateful to the EE department of the Ohio-State University 
for hosting him during the course of this research.
Finally the authors would like to thank the reviewers for their constructive
comments which allowed them to greatly improve the presentation of this research.

\begin{figure}[tbp]
\caption{The inset shows the conduction band profile of the resonant-tunneling structure.
The transmission probability with increased voltage at $T=4.2K$, magnetic field $B=10$ Telsa, and the Landau state $L_0 =0$ for structure A and B. The solid, dotted, and dashed  lines are calculated for $V_D= 0.18, 0.23, 0.28 \; V$ for structure A and $ V_D=0.07, 0.12, 0.17\; V$ for structure B, respectively.}
\label{tvef}
\end{figure}
\begin{figure}[tbp]
\caption{The current-voltage characteristic at $T=4.2 K$ for the structure A and B in presence of magnetic fields. The solid, dotted, dashed, dot-dashed, double dot-dashed lines are for $B=0,5,10,15,20$ Telsa, respectively, for both structures.}
\label{ivst}
\end{figure}
\begin{figure}[tbp]
\caption{The solid line is the shift of onset voltage at $T=4.2 K$ in structure B for increasing magnetic field. The dashed line shows the contribution in Eq.(20) of the energy $1/2 \hbar w_c$ of the  Landau ground (lowest) state.}
\label{vshif}
\end{figure}

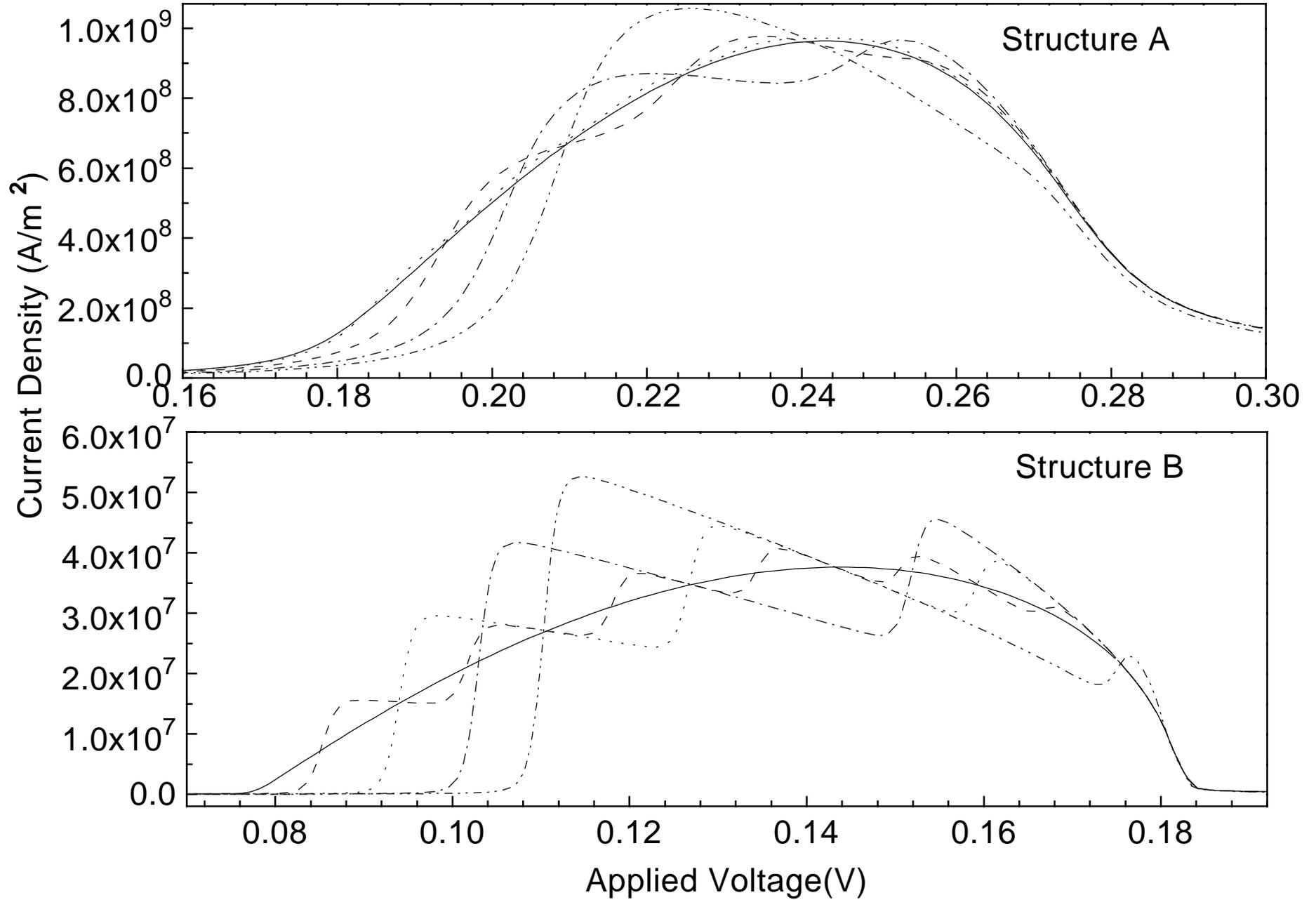
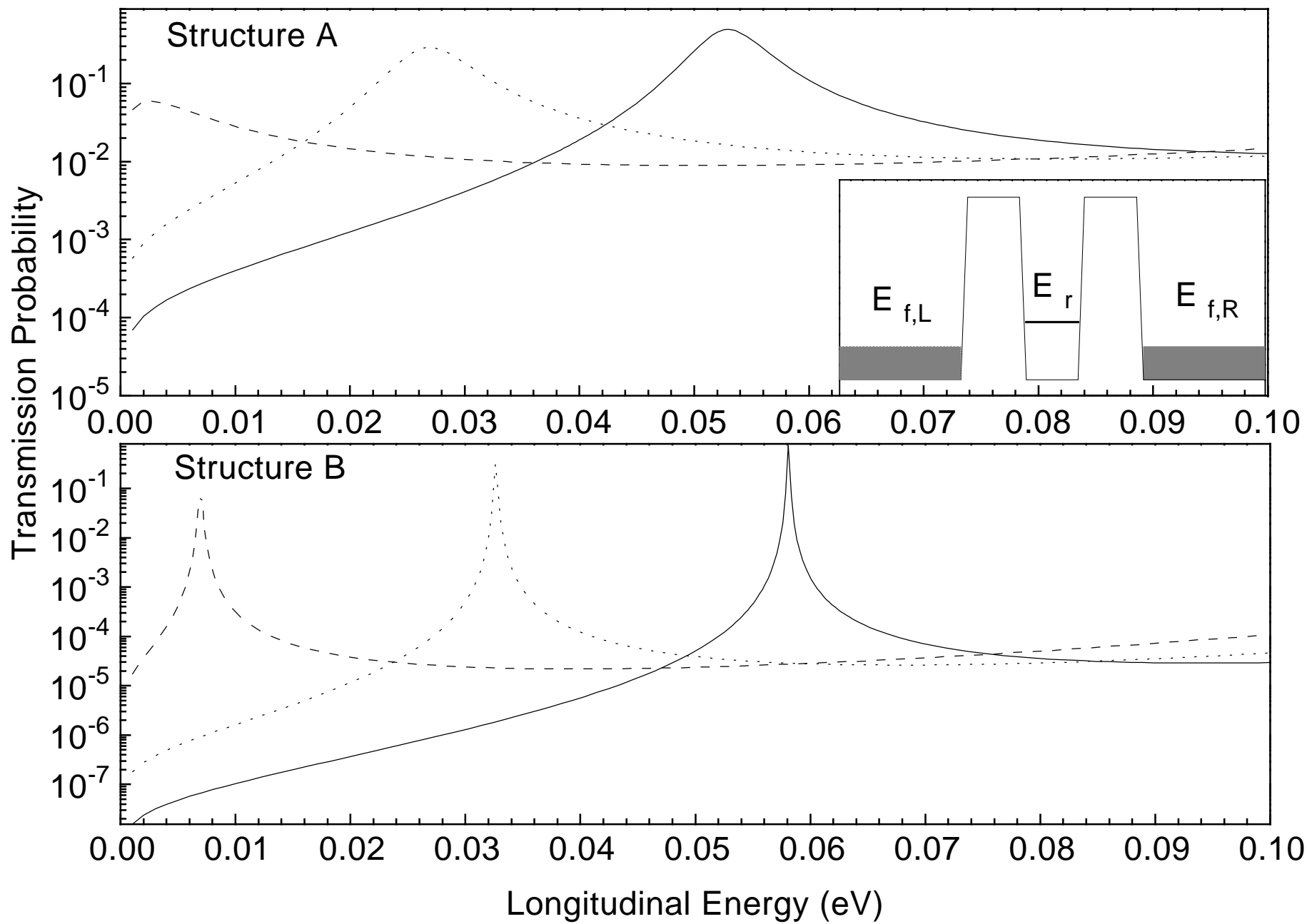
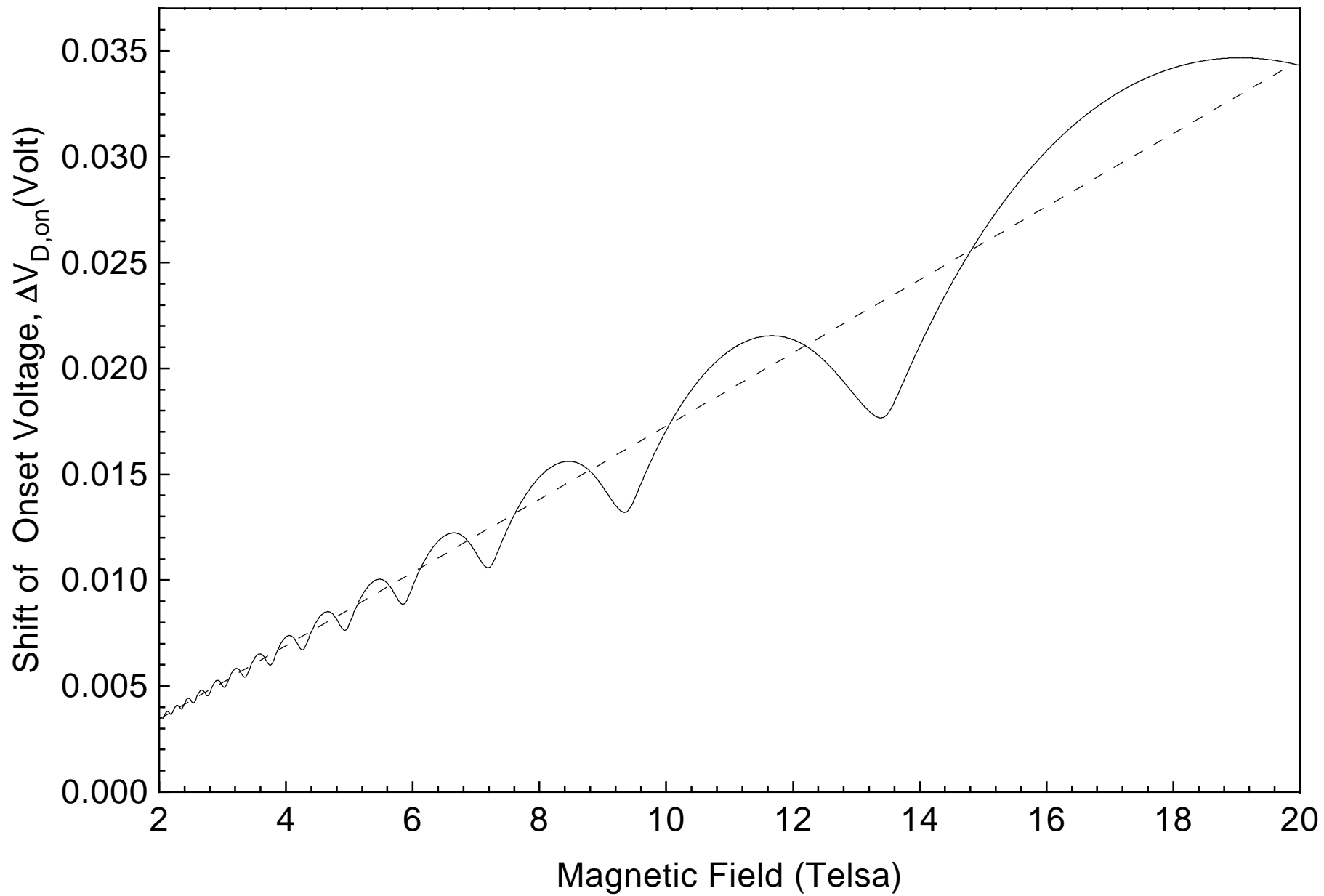
\end{document}